\documentclass[conference]{IEEEtran}
\IEEEoverridecommandlockouts
\usepackage{cite}
\usepackage{amsmath,amssymb,amsfonts}
\usepackage{algorithmic}
\usepackage{algorithm}
\usepackage{graphicx}
\usepackage{textcomp}
\usepackage{xcolor}
\usepackage{gensymb}
\usepackage{subcaption}
\usepackage{multirow}
\usepackage{amsmath}
\usepackage{graphics}
\usepackage{hyperref}
\usepackage[absolute,showboxes]{textpos}

\DeclareMathOperator*{\argmin}{arg\,min}

\TPMargin{5pt}

\newcommand{\copyrightstatement}{
    \begin{textblock}{0.84}(0.08,0.93)    
         \noindent
         \footnotesize
         \copyright  2023 IEEE.  Personal use of this material is permitted.  Permission from IEEE must be obtained for all other uses, in any current or future media, including reprinting/republishing this material for advertising or promotional purposes, creating new collective works, for resale or redistribution to servers or lists, or reuse of any copyrighted component of this work in other works. The peer-reviewed paper is available at \url{https://doi.org/10.1109/IJCNN54540.2023.10191163}.
    \end{textblock}
}

\begin{document}
\copyrightstatement
\bstctlcite{IEEEexample:BSTcontrol}
\title{Morphological Classification of Extragalactic Radio Sources Using Gradient Boosting Methods}


\author{\IEEEauthorblockN{Abdollah Masoud Darya\IEEEauthorrefmark{2}\IEEEauthorrefmark{4}\IEEEauthorrefmark{1},
Ilias Fernini\IEEEauthorrefmark{2},  Marley Vellasco\IEEEauthorrefmark{3}, and
Abir Hussain\IEEEauthorrefmark{4}}
\IEEEauthorblockA{
\IEEEauthorrefmark{2}SAASST, University of Sharjah, Sharjah, UAE\\
\IEEEauthorrefmark{3}Dept. of Electrical Engineering, Pontifical Catholic University of Rio de Janeiro, Rio de Janeiro, Brazil\\
\IEEEauthorrefmark{4}Dept. of Electrical Engineering, University of Sharjah, Sharjah, UAE\\
\IEEEauthorrefmark{1}Corresponding Author: adarya@sharjah.ac.ae}}


\maketitle

\begin{abstract}
The field of radio astronomy is witnessing a boom in the amount of data produced per day due to newly commissioned radio telescopes. One of the most crucial problems in this field is the automatic classification of extragalactic radio sources based on their morphologies. Most recent contributions in the field of morphological classification of extragalactic radio sources have proposed classifiers based on convolutional neural networks. Alternatively, this work proposes gradient boosting machine learning methods accompanied by principal component analysis as data-efficient alternatives to convolutional neural networks. Recent findings have shown the efficacy of gradient boosting methods in outperforming deep learning methods for classification problems with tabular data. The gradient boosting methods considered in this work are based on the XGBoost, LightGBM, and CatBoost implementations. This work also studies the effect of dataset size on classifier performance. A three-class classification problem is considered in this work based on the three main Fanaroff-Riley classes: class 0, class I, and class II, using radio sources from the Best-Heckman sample. All three proposed gradient boosting methods outperformed a state-of-the-art convolutional neural networks-based classifier using less than a quarter of the number of images, with CatBoost having the highest accuracy. This was mainly due to the superior accuracy of gradient boosting methods in classifying Fanaroff-Riley class II sources, with 3--4\% higher recall.
\end{abstract}


\section{Introduction}
Radio observatories such as the upcoming Square Kilometre Array (SKA) aim to improve our understanding of the fundamental phenomena of the universe. To do so, such observatories will generate data in the order of 100s of Terabits per second \cite{dewdney2009square}. Manual classification of the massive amounts of images produced would not be feasible, which calls for an alternative solution.\par

\begin{figure}[tbp]
        \centering
        \begin{subfigure}[b]{0.3\columnwidth}
            \centering
            \includegraphics[width=\textwidth]{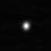}
            \caption{}   
        \end{subfigure}
        \begin{subfigure}[b]{0.3\columnwidth}  
            \centering 
            \includegraphics[width=\textwidth]{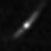}
            \caption{}    
        \end{subfigure}
        \begin{subfigure}[b]{0.3\columnwidth}   
            \centering 
            \includegraphics[width=\textwidth]{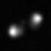}
            \caption{}    
        \end{subfigure}
        \caption{Examples of a) FR0, b) FRI, and c) FRII sources.} 
        \label{fig:1}
\end{figure}

Extragalactic radio sources can be grouped into distinct categories based on their visual appearance. They may appear with either compact or extended morphologies. Most radio sources (around $90\%$) are compact \cite{banfield2015radio}, while extended radio sources can be classified as either Fanaroff \& Riley class I (FRI) or Fanaroff \& Riley class II (FRII). These classifications were first proposed in \cite{fanaroff1974morphology}. FRI and FRII sources can be differentiated by the ratio of the distance between the regions of highest brightness on opposite sides of the central source to the total extent of the source from the lowest contour. The sources for which this ratio is less than $0.5$ are classified as FRI; otherwise, they are considered FRII sources (see Fig. \ref{fig:1} for examples). For simplicity, compact sources will be referred to as Fanaroff \& Riley class 0 (FR0) sources \cite{baldi2019high}.\par

Machine learning has been previously applied to several problems in radio astronomy. For instance, they have been used in identifying the positions of radio sources \cite{lukic2019convosource} and the cross-identification of radio sources with their optical or infrared counterparts \cite{alegre2022machine}. The automated morphological classification of radio sources using deep learning was first proposed in \cite{aniyan2017classifying}. Since then, several studies were also published \cite{alhassan2018first,lukic2018radio,lukic2019morphological,wu2019radio,ma2019machine,becker2021cnn,maslej2021morphological,samudre2022data}.\par

In \cite{samudre2022data}, Samudre \textit{et al.} proposed using DenseNet \cite{huang2017densely}, a transfer-learning-based pre-trained network, as a data-efficient classifier for small labeled datasets ($\approx 2\,000$ images). This work proposes an alternative solution. As opposed to transfer-learning methods, this work proposes gradient boosting methods that require no pre-training and are able to utilize even smaller labeled datasets ($<1\,000$ images) effectively. The performance of the proposed methods is benchmarked with the state-of-the-art Convolutional Neural Network (CNN) proposed in \cite{becker2021cnn}. This comparison is in terms of several standard performance metrics versus dataset size. It is hoped that the findings of this work\footnote{The code and dataset used in this work are available from \\\url{https://github.com/AbdollahMasoud/IJCNN-2023}.} would allow radio astronomers to make an informed decision on whether using CNN or gradient boosting classifiers would be more appropriate, given the size of the dataset at their disposal.\par

\section{Background}
The following subsections present some background information regarding some of the key methods utilized in this work.\par
\subsection{ConvXpress}
In \cite{becker2021cnn}, Becker \textit{et al.} conducted a comparison between the CNN architectures proposed in the literature for the morphological classification of radio sources. They also proposed a novel architecture named ConvXpress (CXP) that is based on the ConvNet-8 \cite{lukic2019morphological}, and VGG-16D \cite{simonyan2014very} architectures. It uses the convolutional stack architecture proposed by VGG-16D and includes linear activation in the final dense layer with L2 kernel regularization, similar to ConvNet-8. The architecture of CXP is presented in Table \ref{tab_cxp}.\par
The CXP model scored the highest in the comparison performed in \cite{becker2021cnn} in terms of balancing between classification and computational performance (see Table 9 in \cite{becker2021cnn}). Therefore, the CXP classifier is used as a benchmark in this work to compare the performance of the three gradient boosting methods. The only modification made to the original CXP architecture was to change the number of output neurons from $4$ to $3$ since Becker \textit{et al.} considered a 4-class classification problem, namely: compact, FRI, FRII, and bent-tailed morphologies, whereas this work considers a 3-class classification problem. The bent-tailed class was excluded from this work as it represents less than $10\%$ of the dataset.\par
The CXP model was trained with a learning rate of $0.001$, a batch size of $64$, a maximum of $30$ epochs, and early stopping with patience of $10$ epochs.\par

\begin{table}[tbp]
\caption{Architecture of the ConvXpress classifier \cite{becker2022application}.}
\begin{center}
\begin{tabular}{|l|c|c|c|c|}
\hline
\textbf{Layer} & \textbf{Depth} & \textbf{Kernel Size} & \textbf{Stride Length} & \textbf{Activation} \\ \hline
Conv2D & $\phantom{0}32$ & $3 \times 3$ & $2$ & ReLU \\ \hline
Conv2D & $\phantom{0}32$ & $3 \times 3$ & $1$ & ReLU \\ \hline
Conv2D & $\phantom{0}32$ & $3 \times 3$ & $1$ & ReLU \\ \hline
MaxPool2D & & $2 \times 2$ & $2$ &  \\ \hline
Dropout &  &  &  &  \\ \hline
Conv2D & $\phantom{0}64$ & $3 \times 3$ & $2$ & ReLU \\ \hline
Conv2D & $\phantom{0}64$ & $3 \times 3$ & $1$ & ReLU \\ \hline
Conv2D & $\phantom{0}64$ & $3 \times 3$ & $1$ & ReLU \\ \hline
MaxPool2D & & $2 \times 2$ & $2$ &  \\ \hline
Dropout &  &  &  &  \\ \hline
Conv2D & $128$ & $3 \times 3$ & $1$ & ReLU \\ \hline
Conv2D & $128$ & $3 \times 3$ & $1$ & ReLU \\ \hline
Conv2D & $128$ & $3 \times 3$ & $1$ & ReLU \\ \hline
MaxPool2D & & $2 \times 2$ & $2$ &  \\ \hline
Dropout &  &  &  &  \\ \hline
Conv2D & $256$ & $3 \times 3$ & $1$ & ReLU \\ \hline
Conv2D & $256$ & $3 \times 3$ & $1$ & ReLU \\ \hline
MaxPool2D & & $2 \times 2$ & $2$ &  \\ \hline
Dropout &  &  &  &  \\ \hline
Flatten &  &  &  &  \\ \hline
Dense & $500$ &  &  & Linear \\ \hline
Dropout &  &  &  &  \\ \hline
Dense & $\phantom{00}3$ &  &  & Softmax \\ \hline
\end{tabular}
\label{tab_cxp}
\end{center}
\end{table}

\subsection{Gradient Boosting Classifiers}
The three methods considered in this work are based on the gradient boosting method proposed by Friedman \cite{friedman2002stochastic}, summarized briefly in this subsection.\par
For a dataset consisting of input variables $\mathbf{x}_i$ and output variable $y_i$ for $i \in [1,N]$, the goal is to find a function $F(\mathbf{x})$ that best maps $\mathbf{x}$ to $y$. To do so, first the gradient boosting model $F(\mathbf{x})$ must be initialized as
\begin{equation}
F_0(\mathbf{x})=\argmin_{\rho} \sum_{i=1}^NL(y_i, \rho),
\end{equation}
where $L(y_i, \rho)$ is the loss function to be minimized. Note that different gradient boosting implementations may use different loss functions. The following steps involve iterating from $m=1$ to the maximum number of trees $M$, where $m$ is the index of each tree. The direction and magnitude in which the loss function can be minimized can be derived from the negative gradient which is defined as
\begin{equation}
    g_{i,m}=-\left[\frac{\partial L(y_i, F(\mathbf{x}_i))}{\partial F(\mathbf{x}_i)}\right]_{F(\mathbf{x})=F_{m-1}(\mathbf{x})}.
\end{equation}
Next, each decision tree is constructed as $T_{l,m}$ for leaves $l \in [1,\mathcal{L}_m]$, where $\mathcal{L}$ is the maximum number of leaves for each tree $m$. Then, a search is conducted for a value of $\rho_{l,m}$ that minimizes the loss function for each leaf
\begin{equation}
    \rho_{l,m} = \argmin_{\rho}\sum_{\mathbf{x}_i\in T_{l,m}}L(y_i, F_{m-1}(\mathbf{x}_i)+\rho).
\end{equation}
The final step is to update the model
\begin{equation}
    F_{m}(\mathbf{x})=F_{m-1}(\mathbf{x})+v\,\rho_{l,m}\,\mathbf{1}(\mathbf{x} \in T_{l,m}),
\end{equation}
where $v$ is the learning rate. The three classifiers considered in this work are based on this framework with some alterations. They are described briefly in the following paragraphs. \par
XGBoost (XGB), short for eXtreme Gradient Boosting, and proposed in \cite{chen2016xgboost} is an efficient open-source implementation of the gradient boosting algorithm. It has several embedded features that distinguish it from other implementations. One of its main benefits over the standard implementation of gradient boosting is its parallelization. It is able to distribute the training process to several CPU cores, considerably improving its training and testing times. It also utilizes L1 and L2 regularization to prevent overfitting.\par
LightGBM (LGB), short for Light Gradient-Boosting Machine, and proposed in \cite{ke2017lightgbm} aimed to improve the efficiency of the gradient boosting method even further. Their proposed algorithm was designed to outperform XGB in terms of scalability and efficiency when the dataset size is large, and the feature dimensions are high. They utilized ``exclusive feature bundling'' that combines mutually exclusive features to reduce the overall number of features. They also used ``gradient-based one-side sampling'' that excludes data with small gradients and uses the rest to estimate the information gain.\par
CatBoost (CAT), short for Categorical Boosting, and proposed in \cite{dorogush2018catboost} was designed to outperform XGB and LGB. To solve the problem of prediction shift that leads to biased models, CAT samples the dataset independently at each boosting step, thus producing unbiased models. CAT also allows categorical features as inputs, which previously had to be converted to numerical values during preprocessing. Another improvement over previous gradient boosting methods is a new scheme to calculate leaf values when constructing the decision tree, considerably reducing overfitting.\par
As the three previously mentioned gradient boosting methods perform exceptionally well for tabular data, the images had to be transformed into a tabular form. This was achieved by flattening all images and concatenating them into a table.\par
\subsection{Principal Component Analysis}
Principal Component Analysis (PCA) was utilized in this work as a feature extraction and dimensionality reduction method. It extracts the principal components representing the highest variance within the dataset, reducing the amount of data to be processed and leading to less memory overhead. Before conducting PCA, the images had to be flattened, and the entire dataset was thus converted to a two-dimensional matrix. More information regarding PCA can be found in \cite{abdi2010principal}.\par

\subsection{Performance Evaluation Metrics}
The performance of the classifiers compared in this work was evaluated in terms of accuracy, precision, recall, and F1-score.
Accuracy can be defined as the number of correct predictions divided by the total number of predictions and is represented by
\begin{equation}
\text{Accuracy}=\frac{\text{TP}+\text{TN}}{\text{TP}+\text{TN}+\text{FP}+\text{FN}},
\end{equation}
where TP, TN, FP, and FN are the number of true positives, true negatives, false positives, and false negatives, respectively. Furthermore, precision and recall are defined as
\begin{equation}
\begin{split}
\text{Precision}&=\frac{\text{TP}}{\text{TP}+\text{FP}},\\
\text{Recall}&=\frac{\text{TP}}{\text{TP}+\text{FN}}.
\end{split}
\end{equation}
The F1-score can be calculated using the precision and recall values and is represented by
\begin{equation}
\text{F1-score}=\frac{2(\text{Precision} \times \text{Recall})}{\text{Precision} + \text{Recall}}.
\end{equation}

\subsection{Dataset}
Similar to \cite{becker2021cnn}, this work utilizes a selection of radio sources from the Best–Heckman sample \cite{best2012fundamental}, manually labeled by \cite{ma2019machine}. The radio images were taken from the Very Large Array Faint Images of the Radio Sky at Twenty Centimeters (FIRST) survey \cite{becker1995first}, and the National Radio Astronomy Observatory Very Large Array Sky Survey (NVSS) \cite{condon1998nrao}. Three classes have been considered in this work. The number of images per class is presented in Table \ref{tab:dataset}. The images were converted from their original FITS format to grey-scale images in JPG format to allow for easier data handling since the JPG images occupy $\approx 200\times$ less storage space than the original FITS files.\par

\begin{table}[tbp]
\caption{Number of images per class.}
\begin{center}
\begin{tabular}{|l|c|c|c|c|}
\hline
\textbf{Class} & FR0 & FRI & FRII & Total\\ \hline
\textbf{Number of Images} & $6\,066$ & $5\,008$ & $2\,066$ & $13\,140$\\\hline
\end{tabular}
\label{tab:dataset}
\end{center}
\end{table}

\section{Methodology}
\begin{figure*}[tbp]
\centerline{\includegraphics[width=\textwidth]{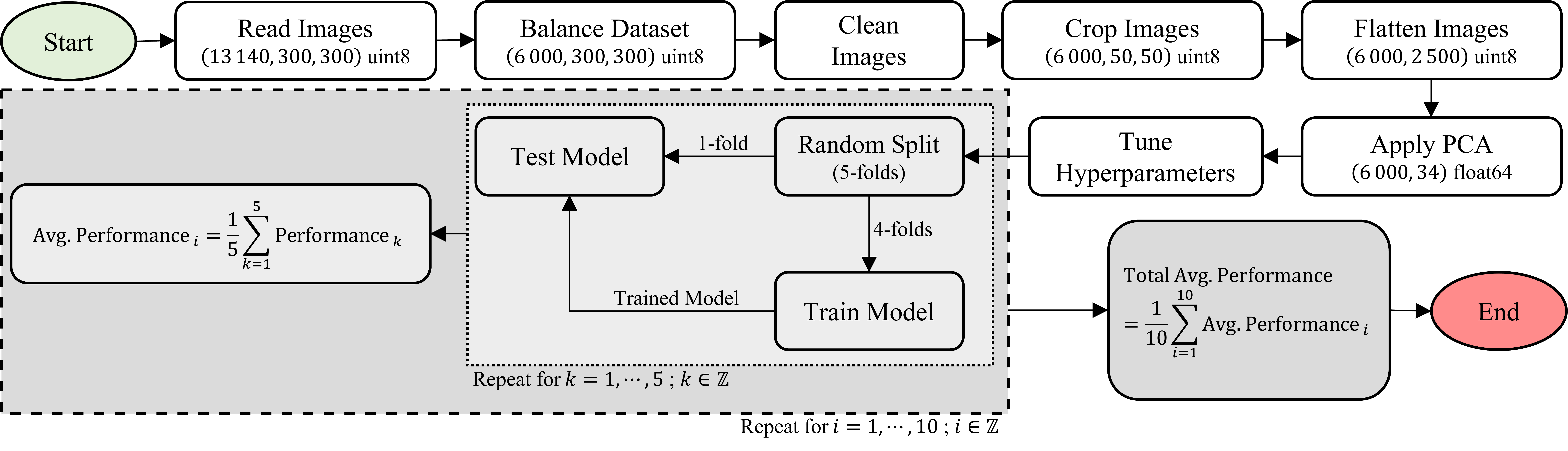}}
\caption{A flowchart of the procedures followed in this work. The shape of the dataset after each process is listed in their corresponding blocks.}
\label{fig:chart}
\end{figure*}

The methodology followed in this work is presented in Fig. \ref{fig:chart} and described in the following subsections.\par

\subsection{Balancing Dataset}
Since the original dataset was not balanced (see Table \ref{tab:dataset}), an important preprocessing step was to balance the dataset to ensure each class was equally represented. It is well known that a machine learning model trained with imbalanced data will be biased towards the class with the higher number of images \cite{brodersen2010balanced}. As the dataset in this work consisted of three classes, the dataset was balanced by taking random samples of $2\,000$ images from each class to form a combined dataset of $6\,000$ images.\par

\subsection{Image Cleaning}
The threshold method used by \cite{becker2021cnn} was applied to reduce the amount of noise caused by radio interference in the images. For each image, all pixels with a magnitude under $\mu+3\sigma$ were set to zero, where $\mu$ is the mean magnitude, and $\sigma$ is the standard deviation. As the source is typically the brightest object in the image, it should remain unchanged, and most of the noise would be removed, thus improving classification performance.\par

\subsection{Image Cropping}
The dimensions of the images under consideration directly influence the overall size of the dataset. By cropping the images to only cover the region of interest, the performance of the classifiers could be improved since this step eliminates radio emissions that are not associated with the main source at the center \cite{lukic2019morphological}. The effect of different cropping sizes on classification performance was studied using the default LGB classifier and the balanced dataset containing $6\,000$ images. The images were cropped from the center, i.e., the center of all resulting images would be the same as the original $300 \times 300$ pixel image. The results of this experiment are presented in Table \ref{tab:imsize}.\par

Table \ref{tab:imsize} shows that the highest classification accuracy and F1-scores of all classes were achieved with a cropped image of size $50 \times 50$ pixels. Furthermore, Table \ref{tab:imsize} shows that a dataset consisting of images of size $50 \times 50$ pixels would be $36\times$ smaller than the original dataset with images of size $300 \times 300$ pixels. Accordingly, in this work, all images were cropped to a size of $50 \times 50$ pixels.\par

\begin{table*}[tbp]
\caption{Performance of the LightGBM classifier versus image size, using unsigned 8-bit integer (uint8) format. The best value for each metric is underlined.}
\begin{center}
\begin{tabular}{|c|c|c|ccc|ccc|ccc|}
\hline
\multirow{2}{*}{\textbf{\begin{tabular}[c]{@{}c@{}}Image Size\\ (pixels)\end{tabular}}} & \multirow{2}{*}{\textbf{\begin{tabular}[c]{@{}c@{}}Dataset Size\\ (per $10^4$ images)\end{tabular}}} & \multirow{2}{*}{\textbf{Accuracy}} & \multicolumn{3}{c|}{\textbf{Precision}} & \multicolumn{3}{c|}{\textbf{Recall}} & \multicolumn{3}{c|}{\textbf{F1-score}} \\ \cline{4-12} 
 &  &  & \multicolumn{1}{c|}{\textbf{FR0}} & \multicolumn{1}{c|}{\textbf{FRI}} & \textbf{FRII} & \multicolumn{1}{c|}{\textbf{FR0}} & \multicolumn{1}{c|}{\textbf{FRI}} & \textbf{FRII} & \multicolumn{1}{c|}{\textbf{FR0}} & \multicolumn{1}{c|}{\textbf{FRI}} & \textbf{FRII} \\ \hline
 $\phantom{0}10\times \phantom{0}10$ & $\phantom{00}1$ MB & $0.7518$ & \multicolumn{1}{c|}{$0.7740$} & \multicolumn{1}{c|}{$0.7121$} & $0.7675$ & \multicolumn{1}{c|}{$0.8044$} & \multicolumn{1}{c|}{$0.6936$} & $0.7574$ & \multicolumn{1}{c|}{$0.7889$} & \multicolumn{1}{c|}{$0.7027$} & $0.7624$ \\ \hline
 $\phantom{0}25\times \phantom{0}25$ & $\phantom{00}6$ MB & $0.7948$ & \multicolumn{1}{c|}{$0.8270$} & \multicolumn{1}{c|}{$0.7525$} & $0.8031$ & \multicolumn{1}{c|}{$0.8520$} & \multicolumn{1}{c|}{\underline{$0.7383$}} & $0.7941$ & \multicolumn{1}{c|}{$0.8393$} & \multicolumn{1}{c|}{$0.7453$} & $0.7986$ \\ \hline
$\phantom{0}50\times \phantom{0}50$ & $\phantom{0}25$ MB & \underline{$0.8039$} & \multicolumn{1}{c|}{\underline{$0.8368$}} & \multicolumn{1}{c|}{\underline{$0.7626$}} & $0.8100$ & \multicolumn{1}{c|}{$0.8617$} & \multicolumn{1}{c|}{$0.7377$} & \underline{$0.8124$} & \multicolumn{1}{c|}{\underline{$0.8490$}} & \multicolumn{1}{c|}{\underline{$0.7500$}} & \underline{$0.8112$} \\ \hline
$100\times 100$ & $100$ MB & $0.8019$ & \multicolumn{1}{c|}{$0.8269$} & \multicolumn{1}{c|}{$0.7612$} & \underline{$0.8148$} & \multicolumn{1}{c|}{\underline{$0.8666$}} & \multicolumn{1}{c|}{$0.7316$} & $0.8075$ & \multicolumn{1}{c|}{$0.8463$} & \multicolumn{1}{c|}{$0.7461$} & $0.8111$ \\ \hline
$150\times 150$ & $225$ MB & $0.7997$ & \multicolumn{1}{c|}{$0.8264$} & \multicolumn{1}{c|}{$0.7558$} & $0.8143$ & \multicolumn{1}{c|}{$0.8645$} & \multicolumn{1}{c|}{$0.7307$} & $0.8040$ & \multicolumn{1}{c|}{$0.8450$} & \multicolumn{1}{c|}{$0.7430$} & $0.8091$ \\ \hline
$200\times 200$ & $400$ MB & $0.7988$ & \multicolumn{1}{c|}{$0.8288$} & \multicolumn{1}{c|}{$0.7545$} & $0.8104$ & \multicolumn{1}{c|}{$0.8618$} & \multicolumn{1}{c|}{$0.7309$} & $0.8036$ & \multicolumn{1}{c|}{$0.8450$} & \multicolumn{1}{c|}{$0.7425$} & $0.8070$ \\ \hline
$250\times 250$ & $625$ MB & $0.7989$ & \multicolumn{1}{c|}{$0.8289$} & \multicolumn{1}{c|}{$0.7554$} & $0.8099$ & \multicolumn{1}{c|}{$0.8613$} & \multicolumn{1}{c|}{$0.7300$} & $0.8056$ & \multicolumn{1}{c|}{$0.8448$} & \multicolumn{1}{c|}{$0.7425$} & $0.8077$ \\ \hline
$300\times 300$ & $900$ MB & $0.7985$ & \multicolumn{1}{c|}{$0.8298$} & \multicolumn{1}{c|}{$0.7542$} & $0.8088$ & \multicolumn{1}{c|}{$0.8601$} & \multicolumn{1}{c|}{$0.7303$} & $0.8050$ & \multicolumn{1}{c|}{$0.8447$} & \multicolumn{1}{c|}{$0.7421$} & $0.8069$ \\ \hline
\end{tabular}
\label{tab:imsize}
\end{center}
\end{table*}

\subsection{Dimensionality Reduction}
\begin{figure}[tbp]
\centerline{\includegraphics[width=\columnwidth]{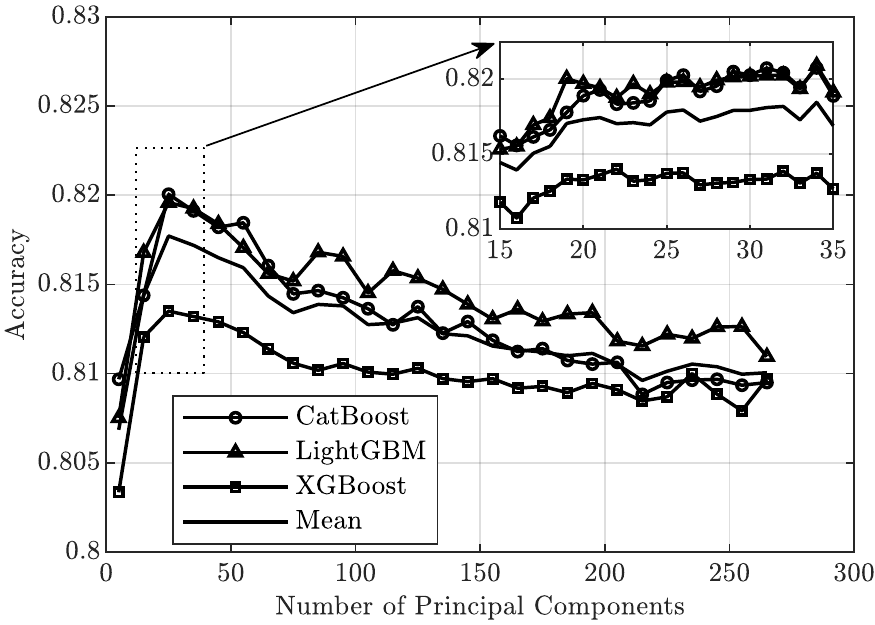}}
\caption{Classifier accuracy versus the number of principal components using the balanced dataset containing $6\,000$ images.}
\label{fig:pca}
\end{figure}

Before conducting PCA, the images had to be flattened from $50 \times 50$ pixels to a variable of size $1 \times 2\,500$ corresponding to a single row of the dataset variable. Each flattened image was then concatenated vertically so that the final dimensions of the transformed dataset were $6\,000 \times 2\,500$, corresponding to $6\,000$ images. Unlike the pixels with magnitude ranging from $0$--$255$ with 8-bit unsigned integer format (uint8), the principal components are represented by 64-bit double-precision floating-point format (float64). Since dimensionality reduction is one of the main aspects of why PCA was used in this work, the size of the transformed data must be equal to or less than its size pre-transformation. As such, the optimal number of principal components was chosen by conducting two searches. First, a broader search was conducted from $5$ to $265$ principal components with a resolution of $10$. Next, another search with a finer resolution of $1$ principal component was conducted from $15$ to $35$. The highest accuracy was achieved with $34$ principal components, as shown in Fig. \ref{fig:pca}.\par

The effect of PCA on classifier accuracy, and dataset size can be seen in Table \ref{tab:pca}. The size of the dataset variable was reduced by $\approx 9\times$ after applying PCA, while the classification accuracy of the three classifiers improved by $\approx 1$--$8\%$. The XGB classifier had the largest improvement in accuracy, especially for a smaller number of images. Since CNN uses convolutional layers for feature extraction, PCA was only applied to the data used to train and test the gradient boosting methods, while the data used to train and test the CXP model was left unchanged. By combining image cropping and PCA, the dataset size was reduced by $\approx 330\times$, as compared to using the original image size and not applying PCA. These methods would be beneficial for handling large numbers of images.\par

\begin{table}[tbp]
\caption{Effect of PCA on dataset size and classifier (Clf.) accuracy.}
\begin{center}
\begin{tabular}{|c|c|c|c|c|c|}
\hline
\textbf{\begin{tabular}[c]{@{}c@{}}Num. of\\Images\end{tabular}} & \textbf{\begin{tabular}[c]{@{}c@{}}Size W/O \\ PCA\end{tabular}} & \textbf{\begin{tabular}[c]{@{}c@{}}Size W \\ PCA\end{tabular}} & \textbf{Clf.} & \textbf{\begin{tabular}[c]{@{}c@{}}Acc. W/O\\PCA\end{tabular}} & \textbf{\begin{tabular}[c]{@{}c@{}}Acc. W\\PCA\end{tabular}} \\ \hline
\multirow{3}{*}{$\phantom{0}300$} & \multirow{3}{*}{$\phantom{0}0.75$ MB} & \multirow{3}{*}{$0.08$ MB} & XGB & $0.6757$ & $0.7540$ \\ \cline{4-6} 
 &  &  & LGB & $0.6853$ & $0.7530$ \\ \cline{4-6} 
 &  &  & CAT & $0.7243$ & $0.7520$ \\ \hline
\multirow{3}{*}{$\phantom{0}600$} & \multirow{3}{*}{$\phantom{0}1.50$ MB} & \multirow{3}{*}{$0.16$ MB} & XGB & $0.7368$ & $0.7847$ \\ \cline{4-6} 
 &  &  & LGB & $0.7418$ & $0.7853$ \\ \cline{4-6} 
 &  &  & CAT & $0.7717$ & $0.7973$ \\ \hline
\multirow{3}{*}{$\phantom{0}900$} & \multirow{3}{*}{$\phantom{0}2.25$ MB} & \multirow{3}{*}{$0.24$ MB} & XGB & $0.7556$ & $0.7972$ \\ \cline{4-6} 
 &  &  & LGB & $0.7666$ & $0.7914$ \\ \cline{4-6} 
 &  &  & CAT & $0.7826$ & $0.8094$ \\ \hline
\multirow{3}{*}{$1\,200$} & \multirow{3}{*}{$\phantom{0}3.00$ MB} & \multirow{3}{*}{$0.32$ MB} & XGB & $0.7625$ & $0.7953$ \\ \cline{4-6} 
 &  &  & LGB & $0.7760$ & $0.7975$ \\ \cline{4-6} 
 &  &  & CAT & $0.7890$ & $0.7998$ \\ \hline
\multirow{3}{*}{$1\,500$} & \multirow{3}{*}{$\phantom{0}3.75$ MB} & \multirow{3}{*}{$0.41$ MB} & XGB & $0.7752$ & $0.7968$ \\ \cline{4-6} 
 &  &  & LGB & $0.7841$ & $0.7957$ \\ \cline{4-6} 
 &  &  & CAT & $0.7863$ & $0.8012$ \\ \hline
\multirow{3}{*}{$3\,000$} & \multirow{3}{*}{$\phantom{0}7.50$ MB} & \multirow{3}{*}{$0.82$ MB} & XGB & $0.7791$ & $0.8079$ \\ \cline{4-6} 
 &  &  & LGB & $0.8008$ & $0.8077$ \\ \cline{4-6} 
 &  &  & CAT & $0.7992$ & $0.8119$ \\ \hline
\multirow{3}{*}{$4\,500$} & \multirow{3}{*}{$11.25$ MB} & \multirow{3}{*}{$1.22$ MB} & XGB & $0.7835$ & $0.8136$ \\ \cline{4-6} 
 &  &  & LGB & $0.8060$ & $0.8156$ \\ \cline{4-6} 
 &  &  & CAT & $0.8016$ & $0.8167$ \\ \hline
\multirow{3}{*}{$6\,000$} & \multirow{3}{*}{$15.00$ MB} & \multirow{3}{*}{$1.63$ MB} & XGB & $0.7793$ & $0.8135$ \\ \cline{4-6} 
 &  &  & LGB & $0.8039$ & $0.8204$ \\ \cline{4-6} 
 &  &  & CAT & $0.8014$ & $0.8203$ \\ \hline
\end{tabular}
\label{tab:pca}
\end{center}
\end{table}

\subsection{Model Tuning}
To obtain the best performance possible from the three gradient boosting classifiers, a grid-search method was used to tune their hyper-parameters. The list of tuned hyperparameters for each classifier is presented in Table \ref{tab:tune}. A description of these hyper-parameters is provided in the next paragraphs.\par
\emph{booster/boosting/boosting\_type}: defines the type of boosting to be utilized. The classic gradient boosting method is defined as \emph{gbtree}, \emph{gbdt}, and \emph{plain} by the XGB, LGB, and CAT classifiers, respectively. The \emph{dart} method, short for Dropouts meet Multiple Additive Regression Trees \cite{vinayak2015dart}, can be selected for the XGB and CAT classifiers. \emph{Dart} is a method that incorporates dropout, a commonly used feature in neural networks, to enhance model regularization. The \emph{gblinear} method can be selected for XGB and utilizes a linear function compared to tree-based models. The Gradient-based One-Side Sampling (GOSS) method, exclusive to LGB,  aims to improve the model's convergence speed by randomly sampling data with small gradients. The \emph{Ordered} method is exclusive to CAT and aims to solve the prediction shift problem by sampling the dataset independently at each boosting step.\par
\emph{learning\_rate}: shrinks the feature weights to prevent overfitting. The CAT method does not have a fixed default \emph{learning\_rate}. Instead, the learning rate is calculated based on the provided dataset. For the utilized dataset the \emph{learning\_rate} calculated by CAT was $0.09$.\par
\emph{max\_depth}: maximum depth of each decision tree. The deeper a tree is, the more complex it is, and the more memory it consumes. While the XGB and CAT methods default to a \emph{max\_depth} $= 6$, LGB defaults to \emph{max\_depth} $= -1$, meaning no limit is placed.\par
\emph{gamma}: used by XGB as a regularization parameter that controls the formation of nodes within a decision tree. The larger the value of \emph{gamma} the higher the regularization. XGB defaults to  \emph{gamma} $= 0$, meaning no regularization is applied.\par
\emph{num\_leaves}: used by LGB to define the maximum number of leaves on one tree. It controls the complexity of the decision trees, as a higher value of \emph{num\_leaves} could lead to overfitting.\par
\emph{l2\_leaf\_reg}: used by CAT to define the coefficient at the L2 regularization term of the cost function. Small values of \emph{l2\_leaf\_reg} may lead to overfitting, while large values of \emph{l2\_leaf\_reg} may lead to underfitting.\par
It is important to note that the performance of the classifiers after tuning their hyperparameters only improved slightly compared to using their default settings (less than $0.5\%$ accuracy improvement). In fact, the default hyperparameters of the LGB classifier were the best-performing set of hyperparameters. Therefore, using the three models with their default settings is acceptable. However, it is important to define the objective of each classifier as `multi-class'. The results presented from this point forth were all produced using the tuned classifiers with the hyperparameters highlighted in bold from Table \ref{tab:tune}.

\begin{table}[tbp]
\caption{List of tuned hyper-parameters. The hyperparameters that resulted in the highest classification accuracy are in bold.}
\begin{center}
\begin{tabular}{|l|l|l|l|}
\hline
\textbf{Classifier} & \textbf{Parameter} & \textbf{\begin{tabular}[c]{@{}l@{}}Default\\ Value\end{tabular}} & \textbf{Grid Search Value} \\ \hline
\multirow{4}{*}{XGBoost} & booster & `gbtree' & \textbf{`gbtree'}, `gblinear', `dart' \\ \cline{2-4} 
 & learning\_rate & 0.3 & 0.025, 0.05, 0.1, 0.2, \textbf{0.3} \\ \cline{2-4} 
 & gamma & 0 & \textbf{0}, 0.1, 0.2, 0.3, 0.4, 0.5 \\ \cline{2-4} 
 & max\_depth & 6 & 0, 3, 6, \textbf{9} \\ \hline
\multirow{4}{*}{LightGBM} & boosting & `gbdt' & \textbf{`gbdt'}, `dart', `goss' \\ \cline{2-4} 
 & learning\_rate & 0.1 & 0.025, 0.05, \textbf{0.1}, 0.2, 0.3 \\ \cline{2-4} 
 & num\_leaves & 31 & 20, \textbf{31}, 40, 50 \\ \cline{2-4} 
 & max\_depth & -1 & \textbf{-1}, 3, 6, 9 \\ \hline
\multirow{4}{*}{CatBoost} & boosting\_type & `Plain' & `Plain', \textbf{`Ordered'} \\ \cline{2-4} 
 & learning\_rate & 0.09 & 0.025, 0.05, \textbf{0.1}, 0.2, 0.3 \\ \cline{2-4} 
 & l2\_leaf\_reg & 3 & 1, \textbf{3}, 6 \\ \cline{2-4} 
 & max\_depth & 6 & 3, \textbf{6}, 9 \\ \hline
\end{tabular}
\label{tab:tune}
\end{center}
\end{table}

\subsection{Splitting \& Augmentation}
Next, the balanced dataset was split into two subsets, one for training and another for testing, with a ratio of $80\%:20\%$, respectively, using the repeated k-fold method, where $k=5$, and the number of repetitions was $10$. The k-fold method randomly generates a $k$ number of splits that cover the entire dataset, and the classifier's performance is averaged over the k-folds. This process was also repeated while generating different splits each time for an even better representation of the performance of the classifiers (see Fig. \ref{fig:chart}), meaning a total of $50$ unique splits were generated, with every $5$ splits covering the entire dataset.\par
Since the CNN-based models benefit considerably from augmented data \cite{maslej2021morphological}, two augmentation steps were followed in this work. For the first, each training image was flipped horizontally and vertically, and for the second step, each image was rotated by $90\degree$, $180\degree$, and $270\degree$. The augmented images were exclusively used to train the CXP model, and no attempt was made to use them for training the proposed gradient boosting classifiers. Table \ref{tab3} presents the total number of images and the splitting ratio between training, validation (only for the CXP model), and testing.\par

\begin{table}[tbp]
\caption{The number of images used for training, validation, and testing. The number of augmented images are in parentheses.}
\begin{center}
\begin{tabular}{|c|c|cc|c|}
\hline
\multirow{2}{*}{\textbf{\begin{tabular}[c]{@{}c@{}}Dataset\\Size (imgs.)\end{tabular}}} & \textbf{XGB, LGB, CAT} & \multicolumn{2}{c|}{\textbf{CXP}} & \multirow{2}{*}{\textbf{Testing}} \\ \cline{2-4}
 & \textbf{Training} & \multicolumn{1}{c|}{\textbf{Training}} & \textbf{Validation} &  \\ \hline
$\phantom{00}\,300$ & $\phantom{0}\,240$ & \multicolumn{1}{c|}{$\phantom{00}\,204$} & $\phantom{0}36$ & $\phantom{00}60$ \\ \hline
$\phantom{00}\,600$ & $\phantom{0}\,480$ & \multicolumn{1}{c|}{$\phantom{00}\,408$} & $\phantom{0}72$ & $\phantom{0}120$ \\ \hline
$\phantom{00}\,900$ & $\phantom{0}\,720$ & \multicolumn{1}{c|}{$\phantom{00}\,612$} & $108$ & $\phantom{0}180$ \\ \hline
$\phantom{0}1\,200$ & $\phantom{0}\,960$ & \multicolumn{1}{c|}{$\phantom{00}\,816$} & $144$ & $\phantom{0}240$ \\ \hline
$\phantom{0}1\,500$ & $1\,200$ & \multicolumn{1}{c|}{$\phantom{0}1\,020$} & $180$ & $\phantom{0}300$ \\ \hline
$\phantom{0}3\,000$ & $2\,400$ & \multicolumn{1}{c|}{$\phantom{0}2\,040$} & $360$ & $\phantom{0}600$ \\ \hline
$\phantom{0}4\,500$ & $3\,600$ & \multicolumn{1}{c|}{$\phantom{0}3\,060$} & $540$ & $\phantom{0}900$ \\ \hline
$\phantom{0}6\,000$ & $4\,800$ & \multicolumn{1}{c|}{$\phantom{0}4\,080$} & $720$ & $1\,200$ \\ \hline
$14\,160$ & - & \multicolumn{1}{c|}{($12\,240$)} & $720$ & $1\,200$ \\ \hline
$26\,400$ & - & \multicolumn{1}{c|}{($24\,480$)} & $720$ & $1\,200$ \\ \hline
\end{tabular}
\label{tab3}
\end{center}
\end{table}

\section{Results \& Discussion}

The performance of the three gradient boosting classifiers and the CXP classifier is listed in Table \ref{tab:comp} in terms of accuracy, precision, recall, and F1-score.\par

\begin{figure}[tbp]
\centerline{\includegraphics[width=\columnwidth]{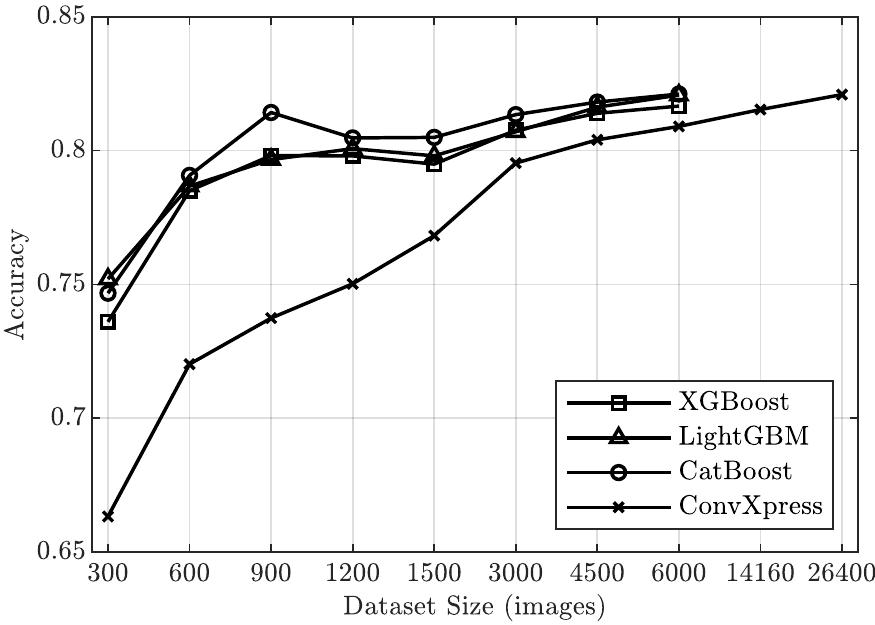}}
\caption{Classifier accuracy versus dataset size.}
\label{fig:comp}
\end{figure}

In Fig. \ref{fig:comp}, the classification accuracy of the three gradient boosting classifiers and the CXP classifier are plotted versus the dataset size defined in terms of the number of images. All three gradient boosting classifiers are seen to outperform the CXP classifier in terms of classification accuracy for dataset sizes ranging from $300$ to $6\,000$ images. Even when the dataset was augmented to a total of $26\,400$ images, the CAT classifier slightly outperformed CXP in terms of accuracy (by $0.02\%$) while using $4.4\times$ fewer images. However, as the dataset size increased, the difference in accuracy between CXP and the three gradient boosting classifiers decreased, from $\approx 8\%$ with $300$ images to $\approx 1\%$ with $6\,000$. This shows that with a higher rate of improvement, the CNN-based CXP classifier benefits more from larger datasets. Therefore, if the dataset was increased to $\approx 50\,000$ images, the CXP classifier is expected to outperform the gradient boosting classifiers.\par

Fig. \ref{fig:comp} also shows that the three gradient boosting methods are very data-efficient, as the CAT classifier exceeded the $80\%$ classification accuracy threshold with a dataset size of $900$ images. Meanwhile, the CXP classifier needed $4\,500$ images ($5\times$ more) to exceed the same threshold. The accuracy of the three gradient boosting classifiers is very similar throughout the trend in Fig. \ref{fig:comp}, with CAT slightly outperforming the other two and the XGB classifier underperforming as compared to the other two. The only difference can be seen at $900$ images, with a noticeable difference between CAT and the other two classifiers.\par

\begin{figure*}[tbp]
        \centering
        \begin{subfigure}[]{\textwidth}
            \centering
            \includegraphics[width=\textwidth]{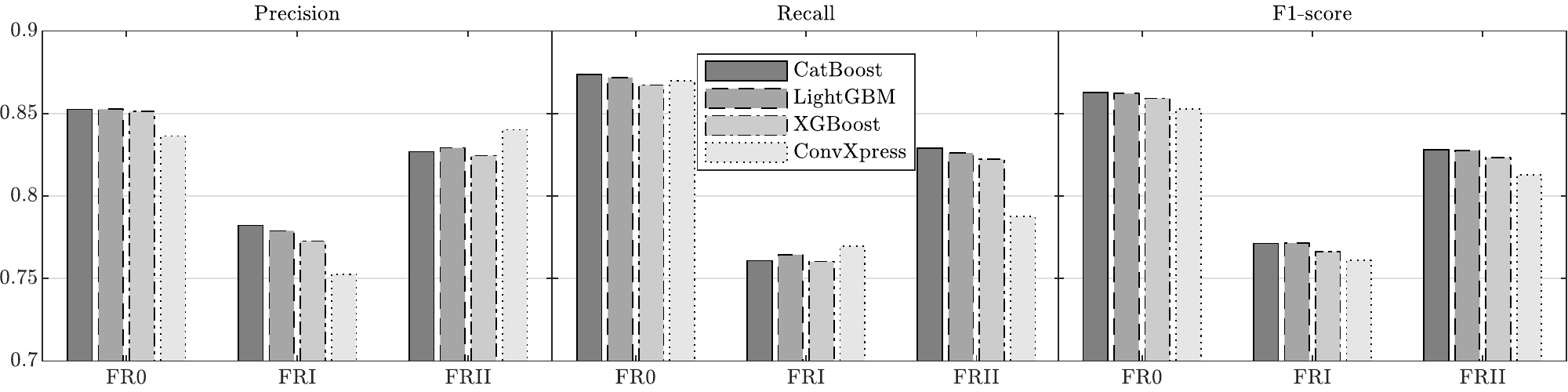}
            \caption{Precision, recall, and F1-score.}   
        \end{subfigure}
        \begin{subfigure}[]{0.245\textwidth}
            \centering
            \includegraphics[width=\textwidth]{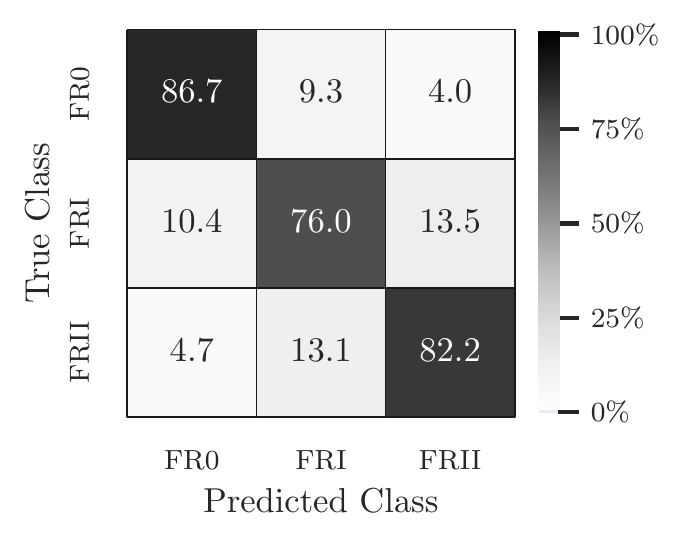}
            \caption{XGB confusion matrix.}   
        \end{subfigure}
        \begin{subfigure}[]{0.245\textwidth}  
            \centering 
            \includegraphics[width=\textwidth]{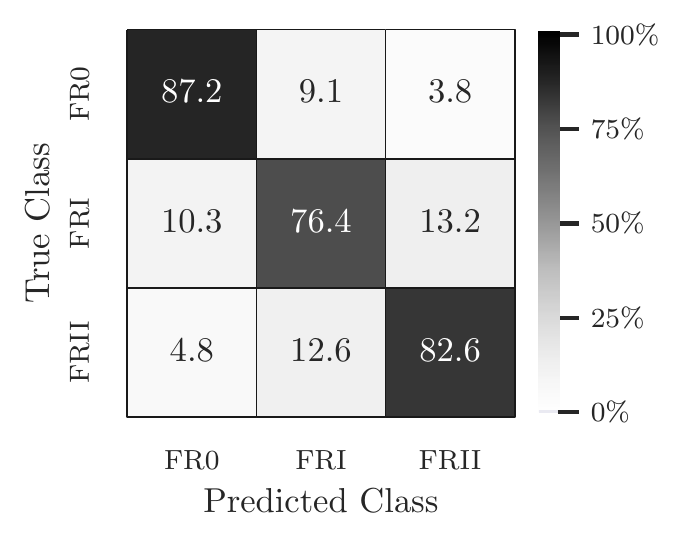}
            \caption{LGB confusion matrix.}    
        \end{subfigure}
        \begin{subfigure}[]{0.245\textwidth}   
            \centering 
            \includegraphics[width=\textwidth]{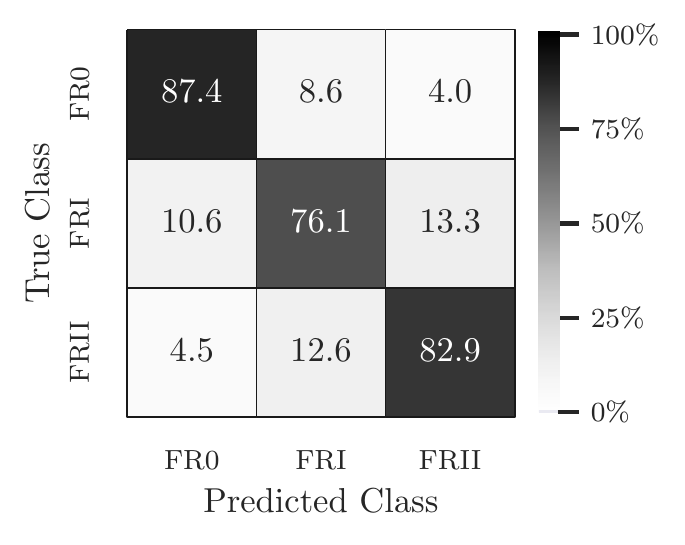}
            \caption{CAT confusion matrix.}    
        \end{subfigure}
        \begin{subfigure}[]{0.245\textwidth}   
            \centering 
            \includegraphics[width=\textwidth]{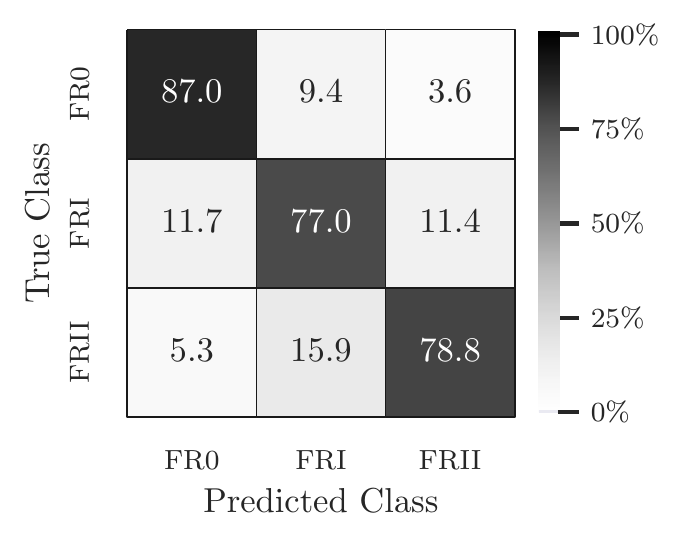}
            \caption{CXP confusion matrix.}    
        \end{subfigure}
        \caption{The performance of all classifiers using the full unaugmented dataset of $6\,000$ images.} 
        \label{fig:CM}
\end{figure*}

The recall, precision, and F1-score of the three gradient boosting classifiers and CXP are presented in Fig. \ref{fig:CM}(a). These evaluation metrics represent the performance of the classifiers using the full unaugmented $6\,000$-image dataset. Out of the three classes, the FR0 class was classified with higher precision, recall, and F1-score than the other classes, i.e., it was classified more accurately. This is mainly due to the compact nature of FR0 sources and their lack of extended radio emissions \cite{baldi2019high}. On the other hand, all classifiers were $\approx 10\%$ less accurate in classifying FRI sources. The confusion matrices in Fig. \ref{fig:CM}(b--e) show the gradient boosting classifiers mislabeling FRI sources as FRII sources $\approx 13\%$ of the time and FR0 sources $\approx 10\%$ of the time. On the other hand, the CXP classifier mislabeled FRI classes evenly between FR0 and FRII sources ($\approx 11\%$ each).\par

When classifying FRII sources, the gradient boosting methods considerably outperformed CXP, with $3.4$--$4.1\%$ higher recall using the XGB and CAT classifiers, respectively. This difference caused the lower classification accuracy of CXP (see Table \ref{tab:comp}) for all dataset sizes. Notably, a FRII recall value above $80\%$ was achieved by all three gradient boosting methods using $900$ images and was not achieved by CXP even with $26\,400$ ($\approx 30\times$ more) images. Therefore, in this work, combining PCA with gradient boosting methods proved superior to CNN for classifying FRII sources.\par
In addition to classification accuracy, it is important to consider the classifiers' training and testing speed. The testing speed, in particular, is important as it indicates the inference speed of the model. A CPU machine with an Intel Xeon 6226R processor and $192$ GB of RAM was used to measure the training and testing time of the classifiers. The findings are presented in Table \ref{tab:time}. The training was conducted using $4\,800$ images ($720$ of which were used for validation by CXP), while testing was conducted using $1\,200$ images, and all times were averaged over $50$ splits (see Fig. \ref{fig:chart}). The training and testing time of all gradient boosting methods was considerably less than the CXP model. Of the three gradient boosting methods, the CAT classifier was the slowest, and the LGB classifier was the fastest. The inference speed of the LGB model was more than $\approx 1\,000 \times$ faster than the CXP model. In fact, for the time it took to test the CXP model, the LGB model could have been trained at least four times over.\par

\begin{table}[tbp]
\caption{Mean training and testing time of all classifiers.}
\begin{center}
\begin{tabular}{|l|c|c|}
\hline
\textbf{Classifier} & \textbf{Training Time (s)} & \textbf{Testing Time (s)} \\ \hline
XGBoost & $\phantom{00}2.6$ & $0.005$ \\ \hline
LightGBM & $\phantom{00}0.6$ & $0.002$ \\ \hline
CatBoost & $\phantom{0}81.0$ & $0.012$ \\ \hline
ConvXpress & $577.6$ & $2.579$ \\ \hline
\end{tabular}
\label{tab:time}
\end{center}
\end{table}

\section{Conclusion}
This work compared the performance of three gradient boosting classifiers based on the CAT, LGB, and XGB implementations with the CNN-based CXP classifier from related work. The three gradient boosting classifiers outperformed CXP in terms of classification accuracy using $4$--$5\times$ fewer images. This was mainly due to the improved accuracy of gradient boosting methods in classifying the FRII class and being more data-efficient. Furthermore, the training and testing speeds of the gradient boosting methods were significantly higher than the CXP model. Most notably, the inference speed of the LGB model was more than $\approx 1\,000\times$ faster than the CXP model. Utilizing PCA as a dimensionality reduction and feature selection method considerably enhanced the performance of the three gradient boosting methods in terms of higher classification accuracy and improved training and testing time while considerably reducing the overall dataset size. By combining the dimensionality reduction effect of PCA, and the image cropping used in this work, the overall dataset size was decreased by $\approx 330\times$ while simultaneously improving the accuracy of the gradient boosting classifiers.\par
The proposed gradient boosting classifiers would be preferable to CNN-based approaches when the number of labeled images is small (sub $10\,000$ images). Additionally, machines with limited computational capabilities can efficiently use the proposed gradient boosting classifiers combined with PCA and image cropping without incurring significant memory/timing overhead. It is hoped that the findings of this work will kindle the interest of the radio astronomy community in gradient boosting methods as data-efficient and computationally affordable alternatives to convolutional neural networks.\par
Potential extensions of this work include exploring explainable gradient boosting methods to understand underlying classification factors \cite{delgado2022implementing}.\par

\begin{table*}[bp]
\caption{Performance of the tuned gradient boosting classifiers and CXP. The best value for each metric is underlined.}
\begin{center}
\resizebox{\textwidth}{!}{%
\begin{tabular}{|c|c|c|ccc|ccc|ccc|}
\hline
\multirow{2}{*}{\textbf{Classifier}} & \multirow{2}{*}{\textbf{\begin{tabular}[c]{@{}c@{}}Dataset Size (images)\end{tabular}}} & \multirow{2}{*}{\textbf{Accuracy}} & \multicolumn{3}{c|}{\textbf{Precision}} & \multicolumn{3}{c|}{\textbf{Recall}} & \multicolumn{3}{c|}{\textbf{F1-score}} \\ \cline{4-12} 
 &  &  & \multicolumn{1}{c|}{\textbf{FR0}} & \multicolumn{1}{c|}{\textbf{FRI}} & \textbf{FRII} & \multicolumn{1}{c|}{\textbf{FR0}} & \multicolumn{1}{c|}{\textbf{FRI}} & \textbf{FRII} & \multicolumn{1}{c|}{\textbf{FR0}} & \multicolumn{1}{c|}{\textbf{FRI}} & \textbf{FRII} \\ \hline
\multirow{8}{*}{XGBoost} & $\phantom{00}\,300$ & $0.7360$ & \multicolumn{1}{c|}{$0.7763$} & \multicolumn{1}{c|}{$0.6853$} & $0.7388$ & \multicolumn{1}{c|}{$0.8190$} & \multicolumn{1}{c|}{$0.6140$} & $0.7750$ & \multicolumn{1}{c|}{$0.7971$} & \multicolumn{1}{c|}{$0.6477$} & $0.7565$ \\ \cline{2-12} 
 & $\phantom{00}\,600$ & $0.7850$ & \multicolumn{1}{c|}{$0.8250$} & \multicolumn{1}{c|}{$0.7219$} & $0.8037$ & \multicolumn{1}{c|}{$0.8720$} & \multicolumn{1}{c|}{$0.6930$} & $0.7900$ & \multicolumn{1}{c|}{$0.8478$} & \multicolumn{1}{c|}{$0.7071$} & $0.7968$ \\ \cline{2-12} 
 & $\phantom{00}\,900$ & $0.7981$ & \multicolumn{1}{c|}{$0.8280$} & \multicolumn{1}{c|}{$0.7525$} & $0.8102$ & \multicolumn{1}{c|}{$0.8603$} & \multicolumn{1}{c|}{$0.7143$} & $0.8197$ & \multicolumn{1}{c|}{$0.8439$} & \multicolumn{1}{c|}{$0.7329$} & $0.8149$ \\ \cline{2-12} 
 & $\phantom{0}1\,200$ & $0.7980$ & \multicolumn{1}{c|}{$0.8315$} & \multicolumn{1}{c|}{$0.7592$} & $0.7994$ & \multicolumn{1}{c|}{$0.8623$} & \multicolumn{1}{c|}{$0.7070$} & $0.8248$ & \multicolumn{1}{c|}{$0.8466$} & \multicolumn{1}{c|}{$0.7322$} & $0.8119$ \\ \cline{2-12} 
 & $\phantom{0}1\,500$ & $0.7949$ & \multicolumn{1}{c|}{$0.8307$} & \multicolumn{1}{c|}{$0.7473$} & $0.8025$ & \multicolumn{1}{c|}{$0.8624$} & \multicolumn{1}{c|}{$0.7034$} & $0.8190$ & \multicolumn{1}{c|}{$0.8462$} & \multicolumn{1}{c|}{$0.7247$} & $0.8107$ \\ \cline{2-12} 
 & $\phantom{0}3\,000$ & $0.8076$ & \multicolumn{1}{c|}{$0.8402$} & \multicolumn{1}{c|}{$0.7650$} & $0.8144$ & \multicolumn{1}{c|}{$0.8721$} & \multicolumn{1}{c|}{$0.7300$} & $0.8207$ & \multicolumn{1}{c|}{$0.8558$} & \multicolumn{1}{c|}{$0.7471$} & $0.8176$ \\ \cline{2-12} 
 & $\phantom{0}4\,500$ & $0.8139$ & \multicolumn{1}{c|}{$0.8448$} & \multicolumn{1}{c|}{$0.7708$} & $0.8241$ & \multicolumn{1}{c|}{$0.8748$} & \multicolumn{1}{c|}{$0.7512$} & $0.8157$ & \multicolumn{1}{c|}{$0.8595$} & \multicolumn{1}{c|}{$0.7609$} & $0.8199$ \\ \cline{2-12} 
 & $\phantom{0}6\,000$ & $0.8166$ & \multicolumn{1}{c|}{$0.8513$} & \multicolumn{1}{c|}{$0.7727$} & $0.8245$ & \multicolumn{1}{c|}{$0.8673$} & \multicolumn{1}{c|}{$0.7603$} & $0.8222$ & \multicolumn{1}{c|}{$0.8592$} & \multicolumn{1}{c|}{$0.7664$} & $0.8233$ \\ \hline
\multirow{8}{*}{LightGBM} & $\phantom{00}\,300$ & $0.7520$ & \multicolumn{1}{c|}{$0.7837$} & \multicolumn{1}{c|}{$0.7000$} & $0.7657$ & \multicolumn{1}{c|}{$0.8440$} & \multicolumn{1}{c|}{$0.6440$} & $0.7680$ & \multicolumn{1}{c|}{$0.8127$} & \multicolumn{1}{c|}{$0.6708$} & $0.7668$ \\ \cline{2-12} 
 & $\phantom{00}\,600$ & $0.7867$ & \multicolumn{1}{c|}{$0.8187$} & \multicolumn{1}{c|}{$0.7305$} & $0.8059$ & \multicolumn{1}{c|}{$0.8785$} & \multicolumn{1}{c|}{$0.6925$} & $0.7890$ & \multicolumn{1}{c|}{$0.8476$} & \multicolumn{1}{c|}{$0.7110$} & $0.7974$ \\ \cline{2-12} 
 & $\phantom{00}\,900$ & $0.7966$ & \multicolumn{1}{c|}{$0.8161$} & \multicolumn{1}{c|}{$0.7525$} & $0.8172$ & \multicolumn{1}{c|}{$0.8710$} & \multicolumn{1}{c|}{$0.7063$} & $0.8123$ & \multicolumn{1}{c|}{$0.8426$} & \multicolumn{1}{c|}{$0.7287$} & $0.8148$ \\ \cline{2-12} 
 & $\phantom{0}1\,200$ & $0.8007$ & \multicolumn{1}{c|}{$0.8271$} & \multicolumn{1}{c|}{$0.7612$} & $0.8097$ & \multicolumn{1}{c|}{$0.8693$} & \multicolumn{1}{c|}{$0.7115$} & $0.8213$ & \multicolumn{1}{c|}{$0.8476$} & \multicolumn{1}{c|}{$0.7355$} & $0.8154$ \\ \cline{2-12} 
 & $\phantom{0}1\,500$ & $0.7980$ & \multicolumn{1}{c|}{$0.8270$} & \multicolumn{1}{c|}{$0.7506$} & $0.8121$ & \multicolumn{1}{c|}{$0.8682$} & \multicolumn{1}{c|}{$0.7072$} & $0.8186$ & \multicolumn{1}{c|}{$0.8471$} & \multicolumn{1}{c|}{$0.7282$} & $0.8153$ \\ \cline{2-12} 
 & $\phantom{0}3\,000$ & $0.8070$ & \multicolumn{1}{c|}{$0.8359$} & \multicolumn{1}{c|}{$0.7618$} & $0.8203$ & \multicolumn{1}{c|}{$0.8711$} & \multicolumn{1}{c|}{$0.7322$} & $0.8177$ & \multicolumn{1}{c|}{$0.8531$} & \multicolumn{1}{c|}{$0.7467$} & $0.8190$ \\ \cline{2-12} 
 & $\phantom{0}4\,500$ & $0.8162$ & \multicolumn{1}{c|}{$0.8458$} & \multicolumn{1}{c|}{$0.7727$} & $0.8281$ & \multicolumn{1}{c|}{$0.8777$} & \multicolumn{1}{c|}{$0.7550$} & $0.8159$ & \multicolumn{1}{c|}{$0.8614$} & \multicolumn{1}{c|}{$0.7638$} & $0.8219$ \\ \cline{2-12} 
 & $\phantom{0}6\,000$ & $0.8207$ & \multicolumn{1}{c|}{\underline{$0.8526$}} & \multicolumn{1}{c|}{$0.7789$} & $0.8292$ & \multicolumn{1}{c|}{$0.8718$} & \multicolumn{1}{c|}{$0.7644$} & $0.8261$ & \multicolumn{1}{c|}{$0.8620$} & \multicolumn{1}{c|}{$0.7716$} & $0.8276$ \\ \hline
\multirow{8}{*}{CatBoost} & $\phantom{00}\,300$ & $0.7467$ & \multicolumn{1}{c|}{$0.7822$} & \multicolumn{1}{c|}{$0.6802$} & $0.7737$ & \multicolumn{1}{c|}{$0.8260$} & \multicolumn{1}{c|}{$0.6550$} & $0.7590$ & \multicolumn{1}{c|}{$0.8035$} & \multicolumn{1}{c|}{$0.6673$} & $0.7663$ \\ \cline{2-12} 
 & $\phantom{00}\,600$ & $0.7907$ & \multicolumn{1}{c|}{$0.8340$} & \multicolumn{1}{c|}{$0.7236$} & $0.8130$ & \multicolumn{1}{c|}{$0.8665$} & \multicolumn{1}{c|}{$0.7185$} & $0.7870$ & \multicolumn{1}{c|}{$0.8499$} & \multicolumn{1}{c|}{$0.7210$} & $0.7998$ \\ \cline{2-12} 
 & $\phantom{00}\,900$ & $0.8142$ & \multicolumn{1}{c|}{$0.8434$} & \multicolumn{1}{c|}{$0.7707$} & $0.8257$ & \multicolumn{1}{c|}{$0.8687$} & \multicolumn{1}{c|}{$0.7373$} & \underline{$0.8367$} & \multicolumn{1}{c|}{$0.8558$} & \multicolumn{1}{c|}{$0.7537$} & \underline{$0.8311$} \\ \cline{2-12} 
 & $\phantom{0}1\,200$ & $0.8047$ & \multicolumn{1}{c|}{$0.8297$} & \multicolumn{1}{c|}{$0.7634$} & $0.8179$ & \multicolumn{1}{c|}{$0.8645$} & \multicolumn{1}{c|}{$0.7278$} & $0.8218$ & \multicolumn{1}{c|}{$0.8467$} & \multicolumn{1}{c|}{$0.7452$} & $0.8198$ \\ \cline{2-12} 
 & $\phantom{0}1\,500$ & $0.8049$ & \multicolumn{1}{c|}{$0.8361$} & \multicolumn{1}{c|}{$0.7613$} & $0.8134$ & \multicolumn{1}{c|}{$0.8716$} & \multicolumn{1}{c|}{$0.7202$} & $0.8228$ & \multicolumn{1}{c|}{$0.8535$} & \multicolumn{1}{c|}{$0.7402$} & $0.8181$ \\ \cline{2-12} 
 & $\phantom{0}3\,000$ & $0.8134$ & \multicolumn{1}{c|}{$0.8419$} & \multicolumn{1}{c|}{$0.7742$} & $0.8208$ & \multicolumn{1}{c|}{$0.8768$} & \multicolumn{1}{c|}{$0.7369$} & $0.8264$ & \multicolumn{1}{c|}{$0.8590$} & \multicolumn{1}{c|}{$0.7551$} & $0.8236$ \\ \cline{2-12} 
 & $\phantom{0}4\,500$ & $0.8181$ & \multicolumn{1}{c|}{$0.8517$} & \multicolumn{1}{c|}{$0.7752$} & $0.8254$ & \multicolumn{1}{c|}{$0.8782$} & \multicolumn{1}{c|}{$0.7579$} & $0.8182$ & \multicolumn{1}{c|}{\underline{$0.8648$}} & \multicolumn{1}{c|}{$0.7665$} & $0.8218$ \\ \cline{2-12} 
 & $\phantom{0}6\,000$ & \underline{$0.8211$} & \multicolumn{1}{c|}{$0.8525$} & \multicolumn{1}{c|}{\underline{$0.7822$}} & $0.8268$ & \multicolumn{1}{c|}{$0.8735$} & \multicolumn{1}{c|}{$0.7608$} & $0.8291$ & \multicolumn{1}{c|}{$0.8629$} & \multicolumn{1}{c|}{$0.7713$} & $0.8279$ \\ \hline
\multirow{10}{*}{ConvXpress} & $\phantom{00}\,300$ & $0.6633$ & \multicolumn{1}{c|}{$0.6758$} & \multicolumn{1}{c|}{$0.5940$} & $0.6999$ & \multicolumn{1}{c|}{\underline{$0.9110$}} & \multicolumn{1}{c|}{$0.4330$} & $0.6460$ & \multicolumn{1}{c|}{$0.7760$} & \multicolumn{1}{c|}{$0.5009$} & $0.6719$ \\ \cline{2-12} 
 & $\phantom{00}\,600$ & $0.7202$ & \multicolumn{1}{c|}{$0.7723$} & \multicolumn{1}{c|}{$0.6249$} & $0.7599$ & \multicolumn{1}{c|}{$0.9055$} & \multicolumn{1}{c|}{$0.6190$} & $0.6360$ & \multicolumn{1}{c|}{$0.8336$} & \multicolumn{1}{c|}{$0.6220$} & $0.6924$ \\ \cline{2-12} 
 & $\phantom{00}\,900$ & $0.7374$ & \multicolumn{1}{c|}{$0.7761$} & \multicolumn{1}{c|}{$0.6806$} & $0.7421$ & \multicolumn{1}{c|}{$0.9010$} & \multicolumn{1}{c|}{$0.5910$} & $0.7203$ & \multicolumn{1}{c|}{$0.8339$} & \multicolumn{1}{c|}{$0.6326$} & $0.7311$ \\ \cline{2-12} 
 & $\phantom{0}1\,200$ & $0.7502$ & \multicolumn{1}{c|}{$0.8338$} & \multicolumn{1}{c|}{$0.6380$} & $0.8006$ & \multicolumn{1}{c|}{$0.8303$} & \multicolumn{1}{c|}{$0.7235$} & $0.6968$ & \multicolumn{1}{c|}{$0.8320$} & \multicolumn{1}{c|}{$0.6781$} & $0.7451$ \\ \cline{2-12} 
 & $\phantom{0}1\,500$ & $0.7682$ & \multicolumn{1}{c|}{$0.8183$} & \multicolumn{1}{c|}{$0.6773$} & $0.8158$ & \multicolumn{1}{c|}{$0.8741$} & \multicolumn{1}{c|}{$0.7125$} & $0.7176$ & \multicolumn{1}{c|}{$0.8452$} & \multicolumn{1}{c|}{$0.6945$} & $0.7635$ \\ \cline{2-12} 
 & $\phantom{0}3\,000$ & $0.7953$ & \multicolumn{1}{c|}{$0.8338$} & \multicolumn{1}{c|}{$0.7203$} & $0.8377$ & \multicolumn{1}{c|}{$0.8629$} & \multicolumn{1}{c|}{$0.7561$} & $0.7669$ & \multicolumn{1}{c|}{$0.8481$} & \multicolumn{1}{c|}{$0.7377$} & $0.8007$ \\ \cline{2-12} 
 & $\phantom{0}4\,500$ & $0.8040$ & \multicolumn{1}{c|}{$0.8436$} & \multicolumn{1}{c|}{$0.7350$} & $0.8390$ & \multicolumn{1}{c|}{$0.8578$} & \multicolumn{1}{c|}{$0.7743$} & $0.7800$ & \multicolumn{1}{c|}{$0.8506$} & \multicolumn{1}{c|}{$0.7541$} & $0.8084$ \\ \cline{2-12} 
 & $\phantom{0}6\,000$ & $0.8090$ & \multicolumn{1}{c|}{$0.8363$} & \multicolumn{1}{c|}{$0.7526$} & $0.8402$ & \multicolumn{1}{c|}{$0.8699$} & \multicolumn{1}{c|}{$0.7697$} & $0.7876$ & \multicolumn{1}{c|}{$0.8528$} & \multicolumn{1}{c|}{$0.7611$} & $0.8130$ \\ \cline{2-12} 
 & $14\,160$ & $0.8153$ & \multicolumn{1}{c|}{$0.8360$} & \multicolumn{1}{c|}{$0.7616$} & $0.8512$ & \multicolumn{1}{c|}{$0.8811$} & \multicolumn{1}{c|}{$0.7793$} & $0.7855$ & \multicolumn{1}{c|}{$0.8579$} & \multicolumn{1}{c|}{$0.7703$} & $0.8170$ \\ \cline{2-12} 
 & $26\,400$ & $0.8209$ & \multicolumn{1}{c|}{$0.8425$} & \multicolumn{1}{c|}{$0.7676$} & \underline{$0.8554$} & \multicolumn{1}{c|}{$0.8802$} & \multicolumn{1}{c|}{\underline{$0.7872$}} & $0.7952$ & \multicolumn{1}{c|}{$0.8609$} & \multicolumn{1}{c|}{\underline{$0.7773$}} & $0.8242$ \\ \hline
\end{tabular}%
}
\label{tab:comp}
\end{center}
\end{table*}

\section*{Acknowledgment}
The authors wish to thank several individuals for their valuable contributions to this work, including Burger Becker for his assistance in reproducing the ConvXpress model, Aisha Al-Owais and Noora Alameri for the fruitful discussions, and Radhia Fernini for copy-editing the manuscript.\par

\bibliographystyle{IEEEtran.bst}
\bibliography{Manuscript.bib}
\end{document}